\documentstyle[12pt]{article}
\author{}
\title{}
\begin{document}

\baselineskip = 8mm plus 0.1 mm minus 0.1 mm

\begin{center}

{\bf STRUCTURES OF GRAVITATIONAL VACUUM AND THEIR ROLE IN THE UNIVERSE}                                                                

\vspace*{1cm}

Vladimir Burdyuzha${}^{1}$, J.A.de Freitas Pacheco${}^{2}$, Grigory Vereshkov${}^{3}$

${}^{1}$ Astro-Space Center, Lebedev Physical Institute, Russian Academy of Sciences,
         Profsoyuznaya 84/32, Moscow 117810, Russia\\

${}^{2}$ Observatoire de la Cote d'Azur, bld.de l Observatoire, 06304 Nice Cedex 4, France\\

${}^{3}$ Department of Physics, Rostov State University, Stachki 194,Rostov/Don 344104, Russia\\

\vspace*{0.3cm}

\end{center}

\baselineskip = 4mm plus 0.1mm minus 0.1 mm

The production of gravitational vacuum defects and their contribution in energy
density of the Universe are discussed. These topological microstructures could
be produced as the result of defect creation of the Universe from "nothing" as
well as the result of the first relativistic phase transition. They must
be isotropically distributed on background of the expanding Universe. After
Universe inflation these microdefects smoothed, stretched and broke up. 
Parts of them have survived and now they are perceived as the structures of
$\Lambda$-term (quintessence) and unclustered dark matter. It is shown that
for phenomenological description of vacuum topological defects of different
dimensions (worm-holes, micromembranes, microstrings and monopoles) the
parametrizational noninvariant members of Wheeler -DeWitt equation can be used. 
The mathematical illustration of these processes may be the spontaneous breaking
of local Lorentz-invariance of quasi-classical equations of gravity. In addition,
3-dimensional topological defects revalues $\Lambda$-term.

\vspace{0.8cm}

INTRODUCTION

\baselineskip = 8mm plus 0.1 mm minus 0.1 mm

Previously cosmology of gravitational vacuum was not practically discussed
although the influence of gravity on a vacuum was considered\cite{B1}.
On the other hand cosmology of other vacua is very often discussed \cite{B2}. 
To be more exact we can say that we shall consider structures of a gravitational
vacuum condensate.
Among fundamental interactions gravity plays the central role as the determinant
of the space-time structure and as the arena of physical reality \cite{B3}.
The question is to find the internal structure of gravitational vacuum starting
from the quantum regime. The quantum regime of gravity has not been satisfactorily
explained although many approaches have been done \cite{B4}.

We present some analogy between known vacuum structures and hypothetical
structures of gravitational vacuum (as it is known condensates of the quark-gluon
type consist of topological structures - instantons). The general representation
about topological defects is: 3-dimensional topological structures ($D = 3$) -
worm holes; 2-dimensional topological structures ($D = 2$) -membranes;
1-dimensional topological structures ($D = 1$) -strings; point defects
(singularities) ($D = 0$) - gas of topological monopoles. The full theory of
vacuum defects is absent although we understand that the presence  of defects
breaks the symmetry of a system. The difficult question is to know the loss of
which symmetry in quantum theory of gravity applies to the presence of
topological defects of gravitational vacuum. Here it is necessary to use the
experience of vacuum physics.

Even now formulation of the problem of the microscopic description of gravitational
properties of vacuum is absent. From general considerations we may propose that
strong fluctuations of topology could take place on the Planck scale. Probably in
these fluctuations stable structures averaging characteristics of which are
constant or slowly changing in time may exist. We have taken into account that
among possible parametrization-noninvariant potentials of Wheeler-DeWitt equation
are ones topological properties of which can be used for macroscopic description
of the gas of topological defects (worm-holes, micromembranes, microstrings and
monopoles). In our opinion, these circumstances allow us to propose a hypothesis
on which the future theory of quantum gravity and the problem of quantum topological
structures and parametrization-noninvariant of Wheeler-DeWitt equation will be
solved in combination. In the frame of this hypothesis based on mathematical
observations and analogies we suggest the mathematical apparatus of systematic
modelling of the superspace metric and the effective potential permitting us to
choose structures which are concerned with a cosmological point of view.

Thus we discuss gravitational vacuum in which energy-momentum characteristics
depends on the radius of a closed Universe. Our arguments are only heuristic:

1. this is the simplest mathematical model; \\
2. our model is synthesized with general property of Wheeler-DeWitt equation
for a closed isotropic Universe. Gauge invariance of classic theory of gravity
has an additional aspect - the parametrizational invariance on the relative choice
of variables on which are put gauge conditions. In classic theory a
parametrization and a gauge are single operations the division of which on
separate steps is conditional. In quantum geometrodynamics (QGD) the situation
is different - the basic equations of QGD are gauge invariant but
parametrization is gauge noninvariant \cite{B5}. Physical consequences of
the parametrizational noninvariance are not general. Authors \cite{B6} have
suggested to reject this problem and to fix the choice of gauge variables by
the concrete way;\\
3. also a conjecture argument has arisen after the introduction in cosmology of
quintessence: a gravitation vacuum condensate may be considered as a possible
factor of the breaking of local vacuum Lorentz-invariance.\\
Of course gravitational vacuum condensate (as well as and its structures) may
arise after the first relativistic phase transition \cite{B7} but in this
article we remain in the frame of Wheeler-DeWitt conceptions on which phase
transitions are absent.

Generally speaking the nature of dependence of vacuum energy from time is good
unknown and the mathematical simulation has discussion character. Although
compensation mechanisms have probably taken place (vacuum energy may be decreased
by jumps in the result of negative contributions during relativistic phase
transitions \cite{B8}). Models of quintessence \cite{B9}
appeal mainly to classical fields (scalar one as example). We shall show that
in quasiclassical approach quintessence can be simulated using mathematical
structures arising naturally in Wheeler-DeWitt quantum geometrodynamics.

BASIC STATEMENTS

Probably the Universe creation is a quantum geometrodynamical transition from
"nothing" (the state of "nothing" has geometry of zero volume (a=0)) in
the state of a closed 3-space of small sizes having some particles and fields.
That is the closed isotropic Universe was created through tunneling process with
a metric:

$$ds^{2} = N^{2} dt^{2} - a^{2}(t) dl^{2} \eqno(1)$$

$$dl^{2} = d\rho^{2} + sin^{2}\rho(d\Theta^{2} + sin^{2}\Theta d\Phi^{2}) \eqno(2)$$

here $N =\sqrt{g_{00}}$ is a gauge variable which is necessary to fix before
the solution of Einstein equations and a tensor energy-momentum(TEM): $T^{\mu}_{\nu} = diag(\epsilon, -p,-p,-p)$
For simplisity we have restricted by the consideration only 3-space, although a
geometrodynamical transition in the state of a closed space of large dimensions
may be also possible. As known in the modern epoch, vacuum energy has overcome the curvature
(it is a dominant component of the Universe) and now the Universe expands
accelerationally: $\Omega_{\Lambda} \sim 0.7; \Omega_{m} \sim 0.3$ and no contradiction
that the Universe was created closed. Besides vacuum energy (if it was positive)
was decreasing by jumps because of negative contributions during its relativistic
phase transitions \cite{B8}. On the first stage we work with the
classic theory in which energy-carriers are described by the hydrodynamical TEM
and which is locally Lorentz-invariant (generally covariant) and which has
the standard $\Lambda$-term. For a closed Universe Einstein equations are:

$$R^{0}_{0} - \frac{1}{2}R \equiv 3(\frac{\dot{a}^{2}}{N^{2}a^{2}} +
\frac{1}{a^{2}}) = \ae (\epsilon_{s} + \Lambda_{0}) \eqno(3)$$

$$R^{k}_{i} - \frac{\delta^{k}_{i}}{2}R \equiv
\delta^{k}_{i} [\frac{1}{N^{2}}(2 \frac{\ddot{a}}{a} -
2 \frac{\dot{N} \dot{a}}{Na} + \frac{\dot{a}^{2}}{a^{2}}) +
\frac{1}{a^{2}}] = \ae \delta^{k}_{i}(-p_{s} + \Lambda_{0}) \eqno(4)$$

$$-R \equiv 6[\frac{1}{N^{2}}(\frac{\ddot{a}}{a} -
\frac{\dot{N} \dot {a}}{Na} +
\frac{\dot{a}^{2}}{a^{2}}) + \frac{1}{a^{2}}] =
\ae(\epsilon_{s} - 3p_{s} + 4 \Lambda_{0}) \eqno(5)$$

Quantum theory of a closed Universe - quantum geometrodynamics\cite{B10}
is based on the Wheeler idea about a superspace. This idea includes the manifold
of all possible geometries of 3-space, matter and field configurations in which
the Universe wave function is defined. Before quantization of classical theory,
it is necessary to impart the form of Lagrange and Hamilton theory with couples
(we are able to quantify only Hamilton theory but the construction of Hamilton theory
precedes the construction of Lagrange one). Also evidently that two variables
must be in Lagrange formulation of the theory:
dynamical variable $a(t)$ relating to an equation of motion and some
Lagrange multiplier $ \lambda = \lambda(t)$ relating to an equation of
couple. From a infinite multiplicity of variants of the inserting of a Lagrange
multiplier we choose the concrete variant related with quantum geometrodynamics.
In quantum geometrodynamics the task arises which has not classical analogy:
it is necessary to formulate the procedure of operators ordering on a generalized
momentum and a coordinate. It is assumed that the procedure of ordering must be
based on the covariance principle of Wheeler-DeWitt equation in Wheeler superspace.
The metric of superspace $ \gamma(a)$ together with a Lagrange multiplier are
inserted for this conception.

The explicit form of function $ \gamma (a)$ is not fixed  but this conception
allows us to understand in which terms the problem of parametrizational
invariance is formulated. The abovementioned program is carried out if:
$N = \frac{\lambda a}{\gamma(a)}$ where $ \lambda$ is a Lagrange multiplier,
$ \gamma(a)$ is the metric of a superspace. Then we have:

$$3\left(\frac{\gamma^{2}{\dot a}^{2}}{\lambda^{2} a^{4}} + \frac{1}{a^{2}}
\right) = \ae \left[ \epsilon_{s}(a) + \Lambda_{0} \right] \eqno(6)$$

$$\frac{\gamma^{2}}{\lambda^{2} a^{2}} \left( 2 \frac{\ddot {a}}{a} -
2 \frac{\dot {\lambda} \dot {a}}{\lambda a} + 2 \frac{\dot {\gamma} \dot {a}}
{\gamma a} - 3 \frac{\dot {a}^{2}}{a^{2}} \right) -
\frac{1}{a^{2}} = \frac{\ae}{3} \left[ \epsilon_{s}(a)+a \cdot
\frac{d \epsilon_{s}(a)}{da}+ \Lambda_{0} \right] \eqno(7)$$

The system of equations  which is mathematically equivalent to these is obtained
by the variational procedure from some effective action. The gravitational
part of this action is the known expression written for a isotropic Universe.

$$S_g=\int \left(\frac{1}{2\ae} R+ \Lambda_{0} \right){\sqrt {-g}}d^4x \eqno(8)$$

Over this expression some standard operations are necessary to perform:
transformation of it in a quadratic form on
generalized velocity $\dot a$ by excluding the total derivative; integration
on volume of a closed Universe $V=2\pi^2a^3$; inserting of parametrizational
of time. The effective action of matter and radiation is added to the received
result using Rubakov-Lapchinsky recept \cite{B11}. This recept is very
simple: the energy density of matter $\epsilon_{s}(a)$ depending on radius of the
Universe in an effective Lagrangian and as $ \Lambda$-term has the status of
an effective potential energy. Therefore, for receiving of right expression it
is necessary to do a replacement $\Lambda_{0}$ to $\Lambda_{0} + \epsilon_{s}(a)$.
The final expression for effective action is:

$$ S_{\gamma} \{a, \lambda\}= \int L_{\gamma} (a, \lambda) dt, \qquad
L_{\gamma} (a, \lambda) = \frac{6 \pi^{2}}{\ae} \frac{1}{\lambda}
\gamma(a) \dot{a}^{2} - \lambda \frac{aU(a)}{\gamma(a)} \eqno(9)$$

where

$$U(a) = \frac{6 \pi^{2}}{\ae} a - 2 \pi^{2} a^{3} [\epsilon_{s} (a) +\Lambda_{0}]$$

is a total effective potential energy  accounting topology of a closed
Universe, standard $\Lambda$-term and matter. The variation of action on
$\lambda (t)$ gives equation:

$$ \frac{\delta S_{\gamma} \{a, \lambda\}}{\delta \lambda}=
\frac{\partial L_{\gamma} (a, \lambda)}{\partial \lambda}=-\frac{2 \pi^{2}}{\ae \gamma}
\left(\frac{3\gamma^2{\dot a}^2 }{\lambda^2}+3a^2-\ae a^4[\epsilon_{s}(a)+
\Lambda_{0} \right) = 0 \eqno(10)$$

which is mathematically equivalent to the equation of the couple. The variation on
dynamical variable $a(t)$ gives the Lagrange equation:

$$\frac{\delta S_{\gamma} \{a, \lambda\}}{\delta a}=-\frac{d}{dt}\frac {\partial
L_{\gamma} (a, \lambda)}{\partial \dot a}+\frac {\partial L_{\gamma} (a,\lambda)}{\partial a}=0 \eqno(11)$$

After some transformations of the Lagrange equation we have:

$$\frac{d}{dt} \frac{\partial L_{\gamma}(a, \lambda)}{\partial \dot{a}}
- \frac {\partial L{\gamma} (a, \lambda)}{\partial a} \equiv$$

$$\frac{6 \pi^{2} \lambda}{\ae \gamma} \left\{ \frac{\gamma^{2}}{\lambda^{2}}
\left( 2 \ddot {a} +2 \frac{\dot {\gamma} \dot{a}}
{\gamma} - 2 \frac{\dot{\lambda} \dot{a}}{\lambda} - 3 \frac{\dot{a}^{2}}{a} \right)
 - a - \frac{\ae a^{3}}{3} (\epsilon_{s}(a) +a \frac{d \epsilon_{s}(a)}{da} +
\Lambda_{0}) + J \right\} = 0 \eqno(12)$$

where

$$J=\frac{2 \pi^{2} \lambda}{\ae \gamma}\cdot \frac{d\ln (a^{-3} \gamma
(a))}{da} \left( \frac{3 \gamma^{2} \dot {a}^{2}}{\lambda^{2}} + 3a^2-\ae
a^4[\epsilon_{s}(a)+\Lambda_{0}]\right)$$

These equations produce a total system of the Lagrange model. Of course,
these equations must be considered in combination. It is easy to see that for $J=0$
the last equation is mathematically equivalent to the combination of Einstein's equations
which have been written before. Thus we have proved that this model gives us 
the Lagrange method of the description of an isotropic Universe, energy-carriers of which
are setted by functions of the scale factor $\epsilon_s(a),\;p_s(a)$. In classic
theory this result has a methodical character only (Einstein's equations in the
initial form are more convenient to work in classic theory). However we want
to transfer these to the quantum geometrodynamics in which Hamilton formulation
is necessary. Hamilton model can be built on basis of Lagrange one. Note,
that introducing of function $\gamma(a)$ is the operation of parametrization.
The index $\gamma$ shows that action and Lagrangian correspond to the definite
parametrization. A Hamiltonian of our system is built on standard rules:

$$H\Phi_{s}=0 \eqno(13)$$

$$H = P \dot{a} - L = \lambda (\frac{\ae}{24 \pi^{2}} \frac{1}{\gamma} P^{2}+
aU(a)) \eqno(14)$$

where $P = \frac{\partial L}{\partial \dot{a}} = \frac{12 \pi^{2}}{\ae}
\frac{\gamma}{a} \dot{a}$ is a generalized momentum. Note that there are also
the parametrization problems of Wheeler-DeWitt theory. The first problem of
the commutation connection is given by the operator $p$ with accuracy to
$\hat{p} =-i \hbar \frac{\partial}{\partial a} +f(a)$. The second problem is
the ordering of operators in the Hamiltonian (this one is created for any
nontrivial function $\gamma(a)$). The partial solution of these problems is
proposed in the frame of hypothesis of covariant differentiation in a curved
space. In the frame of this hypothesis Wheeler-DeWitt equation has the view:

$$-\frac{\ae \hbar^{2}}{24 \pi^{2}} \frac{1}{\sqrt{\gamma}} \frac{d}{da}
\frac{1}{\sqrt{\gamma}} \frac{d \Phi_{s}(a)}{da} + \frac{1}{\gamma}
[ \frac{6 \pi^{2}}{\ae} a^{2} - 2 \pi^{2} a^{4} (\epsilon_{s}(a) + \Lambda_{0})]
\Phi_{s} (a) = 0 \eqno(15)$$

Here we have introduced the quantum index (s) numerating quantum states of
matter and vacuum. The wave function of the Universe satisfies the condition:

$$\int^{\infty}_{0} \sqrt{\gamma} \; da \Phi^{\star}_{s}(a) \Phi^{\star}_{s'}
(a)= \delta (s - s^{'}) \eqno(16)$$

where $\delta(s - s^{'})$ is the delta function or the discrete $\delta$
symbol in dependence  on the concrete properties of Wheeler-DeWitt equation
solutions. Unfortunately the hypothesis of covariant differentiation does not
totally decide parametrizational problems of the theory. For a more evidential
effect of the parametrizational invariance the multiplicative redefinition of
wave function is necessary:

$\Phi_{s} (a) = \gamma^{1/4} \Psi_{s} (a)$

and Wheeler-DeWitt equation is rewritten in the form:

$$(\frac{\ae \hbar}{12 \pi^{2}})^{2} \frac{d^{2} \psi_{s}}{d a^{2}} +
[a^{2} - \frac{a^{4}}{3} (\ae \epsilon_{s} (a) + \ae \Lambda_{0} +
\ae \epsilon_{GVC} (a))] \Psi_{s} (a) = 0 \eqno(17)$$

where

$$\epsilon_{GVC}(a) = \frac{\ae \hbar^{2}}{192 \pi^{2}} \frac{1}{a^{4}}
(\mu^{''} - \frac{1}{4} (\mu^{'})^{2}) \eqno(18)$$

Here we meant  ${'}$ as a derivative of parametric function $\mu(a) =
ln\gamma(a)$ on the scale factor. All parametrizational noninvariant effects are
collected in the function $\epsilon_{GVC} (a)$, which we have named
the density of energy of gravitational vacuum (gravitational vacuum condensate
$(GVC)$). As it is easy to see that parametrizational noninvariant effects are
clean quantum ones, $\epsilon_{GVC} \sim \hbar^{2}$. The parametrization noninvariant
contributions have not the physical generalization if we do not know the physical
nature of their creation. These contributions have arised from nonconservation
of classical symmetry on quantum level. The experience of modern quantum field
theory speaks that a vacuum state rebuilds when a symmetry is broken. For this
reason we have named parametrizational noninvariant contributions as the
density of GVC energy.

However general symmetric arguments  do not have the clear physical connection to
vacuum energy. In this situation the examples from QCD are useful. In quantum
theory classical conform and chiral symmetries do not conserve resulting in the 
appearence of a quark-gluon condensate. Probably concrete vacuum topological
structures exist in a gravitation vacuum and they are the consequence of the
parametrizational noninvariance of quantum geometrodynamics. From general
considerations it is evident that presence in space-time of topological
defects makes it impossible for any continuous transformations of coordinates and time.
That is, concrete properties of defects permitting us to determine the parametrization
of gauge variables.

COSMOLOGICAL APPLICATION

In cosmology this means that properties of topological microscopic defects
inside space on average are isotropic and homogeneous (isotropization
in brane gas cosmology is also a natural consequence of the dynamics \cite{B12}.
They are contained in the function $\mu(a)$. We propose that all topological quantum
defects with $D \ge 1$ have the typical Planck size. On this reason breaking
up defects with a change of their number in variable volume $V = a^{3} (t)$ must
take place. From simple consideration the number of defects in this volume are:

$N_{D} \sim (\frac{a}{l_{pl}})^{D}, \;\;\; l_{pl} = (G \hbar)^{1/2} =
\frac{(\ae \hbar)^{1/2}}{\sqrt{8 \pi}},$  here $c = 1$.

In accordance with these representations we wait untill the energy density of
the system of topological defects contains a constant part corresponding to
worm-holes and also members of type  $1/a^{3}; 1/a^{2}; 1/a$  corresponging
to gas of point defects, micromembranes and microstrings. Besides, the function
$\epsilon_ {GVC}(a)$ must contain additional members describing interactions of
microdefects between each other. Accounted representations correspond to the
next choice of function $\mu(a)$:

$$\mu(a) = c_{0} ln a + c_{1} a + \frac{1}{2} c_{2} a^{2} + \frac{1}{3}
c_{3} a^{3}, \;\;\; c_{i} = const \eqno(19)$$

After this it get easy:

$$\Lambda_{0} + \epsilon_{GVC} (a) = \Lambda_{0} - \frac{\ae \hbar^{2}}
{768 \pi^{2}}\; c^{2}_{3} + \frac{\ae \hbar^{2}}{192 \pi^{2}}
[- \frac{1}{2} \; c_{2} c_{3} \frac{1}{a} - (\frac{1}{4} c^{2}_{2} +
\frac{1}{2} c_{1} c_{3}) \frac{1}{a^{2}} +$$
$$+ (2 c_{3} - \frac{1}{2} c_{1} c_{2} - \frac{1}{2} c_{0} c_{3}) \frac{1}{a^{3}} +
(c_{2} - \frac{1}{4} c^{2}_{1} - \frac{1}{2} c_{0} c_{2}) \frac{1}{a^{4}} -
\frac{1}{2} c_{0} c_{1} \frac{1}{a^{5}} - (c_{0} + \frac{c^{2}_{0}}{4})
\frac{1}{a^{6}}] \eqno(20)$$

The last three members can be interpretated as energy of gravitational
interaction of defects between each other but their discussion is not the case
since quasiclassical dynamics is only right in a region of large a. Note also
that 3-dimensional topological defects revalues $\Lambda$-term. Observed value of
$\Lambda$-term is:

$$\Lambda = \Lambda_{0} - \frac{\ae \hbar^{2}}{768 \pi^{2}} c^{2}_{3} \eqno(21)$$

Probably the term $\frac{1}{a^{3}}$ may be like to dark matter(DM). This gives
the limitation on parameters of function  $\mu(a)$:

$$\frac{1}{3} \; l^{4}_{pl}\; [2 c_{3} (1 - \frac{c_{0}}{a}) - \frac{1}{2} c_{1}
c_{2}] = \ae M \eqno(22)$$

where $M$ is a mass in volume $a^{3}$. As known Wheeler-DeWitt quantum
geometrodynamics is the extrapolation of quantum-theoretical conceptions on the
scale of the Universe as the whole. The initial state of the Universe from QGD
point of view was located in the region of small values of the scale factor in a
minisuperspace. From classical point of view the initial state of the Universe
is a structureless singular state. Here we must postulate the defect creation
of the Universe if it was born from "nothing".  After the release of defects,
probably in our Universe, the stage of quick expansion (inflation) has taken
place. In the result defects are smoothed, stretched and broken up. Some defects
have left and perceived now as $\Lambda$-term (quintessence) and unclustered DM.
Note again, that topological defects of gravitational vacuum may also be
produced after the first relativistic phase transition, but according to the 
ideas of Wheeler-DeWitt, in the frame of which we are, phase transitions are absent.
More fully physics of topological defects arising during phase transitions is
discussed by T.Kibble \cite{B13}.

CONCLUSION

We have taken into account that among parametrizational noninvariant potentials
of Wheeler-DeWitt equation are ones which have macroscopic properties suitable
to macroscopic description of a gas of topological defects (worm-holes, micromembranes,
microstrings and monopoles). This circumstance allows us to propose that in the future
theory of quantum gravity, the problem of topological structures of gravitational
vacuum and parametrizational noninvariance of Wheeler-DeWitt equation will be
solved jointly (authors \cite{B14} even attempted to parametrize an equation of state of
dark energy).
Also we have shown that quasi-classical corrections (which are proportional
$\sim \hbar^{2}$) are completely defined by a superspace metric (that is $\gamma(a)$).
This means that these corrections in the theory of gravity are not entirely
defined by physics of the 4-dim space-time. In the frame of quantum geometrodynamics
a part of the corrections having an influence on the evolution of the Universe in 4-dim
is defined by physics of a superspace (that is they come from another level). If
we interpret these corrections as defects then this means that defects appear
as the result of the interaction of universes in this superspace. The development of
these ideas may be realized in the frame of tertiary quatization.
Besides, it has been noted that the property of Lorentz invariance attributed to 4-manifold
joints 1-dim time and 3-dim space. Here 3-dim defects (worm-holes) give the contribution
in the Lorentz-invariant $\Lambda$-term). Quantum topological defects with D=0,1,2
give Lorentz-noninvariant contributions in vacuum TEM. Thus we emphasize that
topological defects of gravitational vacuum are quantum structures produced at
the Planck epoch of the evolution of the Universe (Lorentz invariance at the Planck scale
must probably be modified \cite{B15}). Topological defects of the gravitational
vacuum shall be included in the composition of $\Lambda$-term (quintessence) and
unclustered dark matter. Probably, this data allows us to improve our
understanding of the content of the Universe's main components.

\end{document}